\documentclass[aps,prl,preprintnumbers,showpacs,twocolumn,nofootinbib]{revtex4} 
\usepackage{graphicx}
\usepackage{dcolumn}
\usepackage{bm}
\def\lsim{\raisebox{-.4ex}{$\stackrel{<}{\scriptstyle \sim}$\,}}
\def\gsim{\raisebox{-.4ex}{$\stackrel{>}{\scriptstyle \sim}$\,}}

\begin{document}

\preprint{UCI-TR-2005-28}  \preprint{hep-ph/0507066}

\input epsf \renewcommand{\topfraction}{0.8}

\title{Dark matter clues in the muon anomalous magnetic moment}

\author{J.~A.~R.~Cembranos$^*$}
\author{A.~Dobado$^{\dag}$}%
\author{A.~L.~Maroto$^{\dag}$}
\affiliation{$^*$Department of Physics and Astronomy,
 University of California, Irvine, CA 92697 USA\\
$^{\dag}$Departamento de  F\'{\i}sica Te\'orica,
 Universidad Complutense de
  Madrid, 28040 Madrid, Spain}

\date{\today}

\begin{abstract}
We study the possibility to explain the non-baryonic dark matter 
abundance and improve the present fits on the muon anomalous magnetic 
moment through the same new physics. The only  viable way
to solve simultaneously both problems which is known to date is 
by using supersymmetric theories. However in this work we show that
massive brane fluctuations (branons) in large extra-dimensions models
 can provide a more economical alternative to supersymmetry. This is so
because the low-energy 
branon physics depends effectively on only three parameters.
Next collider experiments, such as LHC or ILC, will be sensitive to  
branon phenomenology   in the natural parameter region 
 where the theory is able to account for 
the two effects.
\end{abstract}


\maketitle


The existence of dark matter (DM) is a one of the long-standing problems
in astrophyiscs and cosmology, dating back to the early thirties
when F. Zwicky observed for the first time that the total and
visible masses of rich galaxies disagree in a factor 10-100. Since
then, additional evidence has been obtained from galaxy rotation curves, 
galaxy motions in clusters and,  more recently,  
by 
precise measurements of 
 the temperature fluctuations of the Cosmic Microwave Background (CMB) 
radiation \cite{CMBR}, 
Type Ia supernovae \cite{SNeIa}, large scale distribution of galaxies  
\cite{LSS} or Ly$\alpha$ clouds \cite{Lyalpha}. 

The fact that dark matter cannot be made of any of the 
known particles is one of the most pressing arguments for the
existence of new physics, be it in the
form of new particles or as a modificaction of gravity at large 
distances.  The most favoured particle candidate to account for the 
DM energy density is a 
Weakly Interacting Massive Particle (WIMP) which can provide with the 
non-baryonic DM abundance $\Omega_{\rm NBDM}h^2=0.129 - 0.095$ 
measured by WMAP \cite{CMBR}, in the form of a standard thermal relic. 
The 
particle density is in approximate thermal equilibrium until
 $T\sim M/20$, where  $M$ is the mass of the WIMP which we are
equaling to the scale of New Physics (NP),  $\Lambda_{\rm NP}\sim M$.
 WIMPs thin out by annihilation until
their relic density freezes out when the annihilation rate equals the 
Hubble expansion
rate, $n_{\rm Wimp}\langle \sigma_{\rm A}v\rangle\sim H$. If we assume a 
typical annihilation cross section 
$\sigma_{\rm A}\sim \alpha^2/\Lambda_{\rm NP}^2$
 (where $\alpha$ is the electromagnetic coupling constant), the present 
abundance 
 can be roughly estimated to be:
\begin{equation}
\Omega_{\,\rm Wimp}\sim\left(\frac{\Lambda_{\rm NP}}{100\,{\rm GeV}}\right)^2\,.
\end{equation}
The interesting feature of this result is that the NP which is 
able to explain the
missing matter problem ($\Omega_{\,\rm Wimp}\sim 0.1$),
 could be related with the electroweak sector 
($\Lambda_{\rm NP}\sim 100\,{\rm GeV}$) and 
be accessible in the next generation of collider experiments.
The most popular WIMP candidate is the stable lightest supersymmetric
particle which typically corresponds to a neutralino \cite{neutralino}. 

On the other hand, the success of the Standard Model ({SM}) of 
particles and interactions
has been tested in many different experiments without finding 
very important discrepancies so far. 
A very remarkable example  is  the 
electron magnetic moment:
$\vec{\mu}_e=g_e(e/(2 m_e))\vec{s}$, 
whose gyromagnetic ratio deviates from the value $g_e=2$, 
given by the Dirac equation, as  predicted by quantum 
radiative corrections. This fact has been tested up to a 
relative precision of 0.03 parts per million (ppm) and confirms 
 Quantum Electrodynamics (QED) as the most precise physical
 theory \cite{PDG}.

Curiously one of the most interesting deviation from the 
SM prediction is provided by the muon magnetic moment. 
Indeed $a_\mu=(g_\mu-2)/2$ is not only more sensitive to 
strong and weak interactions than the electron moment, 
but also to NP. The  821 
Collaboration at the
Brookhaven Alternating Gradient Synchrotron has reached a precision 
of 0.5 ppm in the measurement
of such a parameter  \cite{BNL}. 
Taking
 into account  $e^+e^-$ collisions data in order to calculate 
the $\pi^+\pi^-$
 spectral functions, the deviation with respect to the SM prediction is at
 $2.6\,\sigma$ \cite{gm2}: $\delta a_\mu \equiv a_\mu
({\rm exp}) - a_\mu ({\rm SM}) =(23.4 \pm 9.1)\times 10^{-10}$. 
On the other hand, the contribution of NP
to this  parameter can be written generically as:
\begin{equation}
\delta a_\mu=k \times \left(\frac{m_\mu}{\Lambda_{\rm NP}}\right)^2\,,
\end{equation}
where the order of magnitude of the constant 
$k$ depends on the particular model under consideration. 
Notice that in order to be able to explain the current 
discrepancy ($\delta a_\mu\sim 10^{-9}$) with the same NP as
for dark matter, i.e. $\Lambda_{NP}\sim 100$ GeV,
 we should have $k\sim 10^{-3}$. This is again the case for
 some particular supersymmetric models in which the deviation mainly 
comes from new loop diagrams containing neutralinos, charginos and 
sleptons. Thus, if there is really new physics in the
Brookhaven results, supersymmetry would be so far the only known 
theory in
which the two problems, dark matter and the muon anomaly, could be
solved simultaneously \cite{Baltz:2001ts}.
In this work, however, we point out that the brane-world scenario,
originally proposed as an alternative to supersymmetry in the
context of the gauge hierarchy problem \cite{ADD}, 
is also a viable alternative here.

It has been recently found that massive brane fluctutations (branons)
are natural candidates to dark matter in brane-world models
with low tension \cite{CDM,M}. Branon physics can be described at
low energies by an effective action which depends essentially on 
only three parameters: the branon mass $M$, the brane tension scale 
$f$ and the cut-off $\Lambda$ which sets the range of validity of the 
effective theory. This should
be compared with the more than one hundred free parameters of the
Minimal Supersymmetric Standard Model (MSSM) or the five parameters
of the constrained MSSM. 

\begin{table}[bt]
\centering \small{
\begin{tabular}{||c|cccc||}
\hline Experiment
&
$\sqrt{s}$(TeV)& ${\mathcal
L}$(pb$^{-1}$)&$f_0$(GeV)&$M_0$(GeV)\\
\hline
%
%
HERA$^{\,1}
$& 0.3 & 110 &  16 & 152
\\
Tevatron-I$^{\,1}
$& 1.8 & 78 &   157 & 822
\\
Tevatron-I$^{\,2}
$ & 1.8 & 87 &  148 & 872
\\
LEP-II$^{\,2}
$& 0.2 & 600 &  180 & 103
\\
\hline
Tevatron-II$^{\,1}
$& 2.0 & $10^3$ &  256 & 902
\\
Tevatron-II$^{\,2}
$& 2.0 & $10^3$ &   240 & 952
\\
ILC$^{\,2}
$& 0.5 & $2\times 10^5$ &  400 & 250
\\
ILC$^{\,2}
$& 1.0 & $10^6$ &  760 & 500
\\
LHC$^{\,1}
$& 14 & $10^5$ &  1075 & 6481
\\
LHC$^{\,2}
$& 14 & $10^5$ &   797 & 6781
\\
CLIC$^{\,2}
$& 5 & $10^6$ &  2640 & 2500
\\
\hline
\end{tabular}
} \caption{\footnotesize{Limits from 
direct branon searches in colliders (results at the $95\;\%$ c.l.). 
Upper indices $^{1,2}$ denote 
monojet and single photon channels respectively. 
The results for 
HERA, LEP-II and Tevatron run I have been obtained
from real data, whereas those for Tevatron run II, 
ILC, LHC and CLIC  are estimations.
$\sqrt{s}$ is the center of mass energy of the total
process; ${\mathcal L}$ is the total integrated luminosity;
 $f_0$ is the bound on the brane tension scale for one
massless branon ($N=1$) and $M_0$ is the limit on the branon mass for 
small tension $f\rightarrow0$ (see \cite{ACDM} for details).}}
\label{tabHad}
\end{table}

In order to have a sensible effective theory the following
hierarchy among the three parameters is expected $M\,\lsim\,f\,\lsim\,\Lambda$. From
 the point of view of the four dimensional 
effective phenomenology, the massive branons are new 
pseudoscalar fields which can be understood as the
pseudo-Goldstone  bosons corresponding to the spontaneous 
breaking of translational
 invariance in the bulk space produced by the presence of the 
brane \cite{DoMa}. They are
 stable due to  parity invariance on the brane. 
The  SM-branon low-energy effective Lagrangian 
\cite{BSky,ACDM}  can be written as:
\begin{eqnarray}
{\mathcal L}_{Br}&=&
\frac{1}{2}g^{\mu\nu}\partial_{\mu}\pi^\alpha
\partial_{\nu}\pi^\alpha-\frac{1}{2}M^2\pi^\alpha\pi^\alpha
\nonumber  \\
&+&
\frac{1}{8f^4}(4\partial_{\mu}\pi^\alpha
\partial_{\nu}\pi^\alpha-M^2\pi^\alpha\pi^\alpha g_{\mu\nu})
T^{\mu\nu}
\,.\label{lag}
\end{eqnarray}
where $\alpha=1\dots N$, with $N$ the number of branon
species. 
We see that branons interact by pairs with the SM
energy-momentum tensor $T^{\mu\nu}$, and that the coupling
is suppressed by the brane tension $f^4$. 
Limits on the model parameter from tree-level processes in colliders  
are briefly summarized in Table \ref{tabHad}, where one can find not 
only the present restrictions coming from HERA, Tevatron and LEP-II,  
but also the
prospects for future colliders such as ILC, LHC or CLIC \cite{ACDM,CrSt}.
Additional bounds from astrophysics and 
cosmology can be found in \cite{CDM1}.

\begin{figure}[bt]
\begin{center}
\resizebox{6cm}{6.5cm} 
{\includegraphics{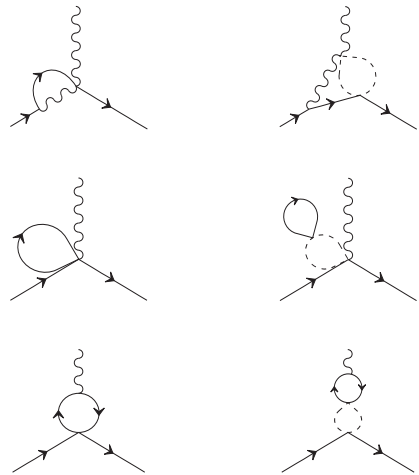}}
\caption{\footnotesize The three diagrams on the left are the different types of
contributions from Lagrangian (\ref{eff}) to the anomalous
magnetic moment of the muon at one loop. 
The three diagrams on the right are the equivalent two loop 
contributions from the branon theory described by (\ref{lag}).
The continuous, dashed and wavy lines represent the muon, 
branon and photon fields respectively. 
Notice that in the first type of contribution, the fermion
loop can also be attached to the outgoing muon.} 
\label{diagrams}
\end{center}
\end{figure}

In order to obtain the first branon contribution to the 
$\mu$ anomalous magnetic moment, we compute the one-loop
effective action for SM particles, by integrating out the
branon fields with cutoff regularized integrals.  At the level of
two-point functions, branon loops result only in a renormalization
of the SM particle masses, which is not observable. However new
couplings appear at higher-point functions which can be described by an effective
Lagrangian \cite{CrSt,radcorr} whose more relevant terms are:
\begin{eqnarray}\label{eff}
{\mathcal L}_{SM}^{(1)}\simeq \frac{N \Lambda^4}{192(4\pi)^2f^8}
\left\{2T_{\mu\nu}T^{\mu\nu}+T_\mu^\mu
T_\nu^\nu\right\}\,.
\end{eqnarray}
As we have commented above, $\Lambda$ is the cutoff which limits the
validity of the effective description of branon and SM dynamics.
This new parameter appears when dealing with branon
radiative corrections since the Lagrangian in (\ref{lag}) is not
renormalizable.
A one-loop calculation with the new effective four-fermion vertices coming
from (\ref{eff}), whose Feynman digrams 
are given in Figure \ref{diagrams},  is  
equivalent to a two-loop computation with the Lagrangian in (\ref{lag}), and
allows us to obtain the contribution of branons 
to the anomalous magnetic moment:
\begin{equation}\label{gb}
\delta a_\mu \simeq \frac{5\, m_\mu^2}{114\,(4\pi)^4}
  \frac{N\Lambda^6}{f^8}.
\end{equation}
This result is qualitatively similar to other 
$\,g_\mu-2\,\,$ contributions obtained in
different analyses in the brane-world scenario
\cite{branemuon,CPRS}.
We can observe that the correction has the correct sign and that
 it is thus possible to
 improve the agreement with the observed experimental 
value by the E821 Collaboration. In fact, 
by using the commented difference between the experimental and 
the SM prediction \cite{BNL,gm2},
 we can estimate the {\it preferred} parameter region for  
branon physics:
\begin{equation}\label{g-2}
    6.0\; \mbox{GeV}\;\gsim\,\frac{f^4}{ N^{1/2}\Lambda^3}\,
\gsim\;2.2\; \mbox{GeV}\; (\,95\; \%\; c.l.\,)
\end{equation}
However branon loops can have additional effects which should also be
compatible with SM phenomenology.  The most relevant ones 
could be the four-fermion interactions or the fermion pair annihilation
 into two gauge bosons. Following  \cite{GS,Hewett},
 we have used the data
coming from HERA \cite{Adloff:2003jm}, Tevatron \cite{d0} 
and LEP \cite{unknown:2004qh}  on this kind of processes in order to
set bounds on the parameter combination $f^2/(\Lambda N^{1/4})$.
The results are shown in Table \ref{radcoll}, where it is also possible 
to find the prospects for  the future colliders mentioned above.
These limits show that an important consequence of 
the relation (\ref{g-2}) is that the first 
branon signals at colliders would 
be associated to radiative corrections  \cite{radcorr} 
and not to the direct 
production studied in previous works \cite{ACDM}.

\begin{table}[bt]
\centering
\begin{tabular}{||c|c c c||} \hline
Experiment          & $\sqrt s$ (TeV) & ${\cal L}$ (pb$^{-1}$) & $f^2/(N^{1/4}\Lambda)$ (GeV) \\ \hline
HERA$^{\,c}$        & 0.3             &  117                   & 52                           \\
Tevatron-I$^{\,a,\,b}$   & 1.8        &  127                   & 69                           \\
LEP-II$^{\,a}$      & 0.2             &  700                   & 59                           \\
LEP-II$^{\,b}$      & 0.2             &  700                   & 75                           \\ \hline
Tevatron-II$^{\,a,\,b}$  & 2.0        & $2 \times 10^3$        & 83                           \\ 
ILC$^{\,b}$         & 0.5             & $5\times 10^5$         & 261                          \\
ILC$^{\,b}$         & 1.0             & $2\times 10^5$         & 421                          \\ 
LHC$^{\,b}$     & 14              & $10^5$                 & 383                          \\ \hline
\end{tabular}
\caption{\label{radcoll} \footnotesize{
Limits from 
virtual branon searches at colliders (results 
at the $95\;\%$ c.l.) The indices $^{a,b,c}$ denote the two-photon,
$e^+e^-$ and $e^+p$ ($e^-p$) channels respectively. 
The first three analysis have been 
performed with real data:
HERA \cite{Adloff:2003jm}, LEP-II \cite{unknown:2004qh} and 
Tevatron-I \cite{d0}; whereas the final four are estimations. 
The first two columns are the same as in Table \ref{tabHad}, 
and the third one corresponds to the lower bound on 
$f^2/(N^{1/4}\Lambda)$.}}
\end{table}

\begin{figure}[bt]
\begin{center}
\resizebox{8cm}{!}
{\includegraphics{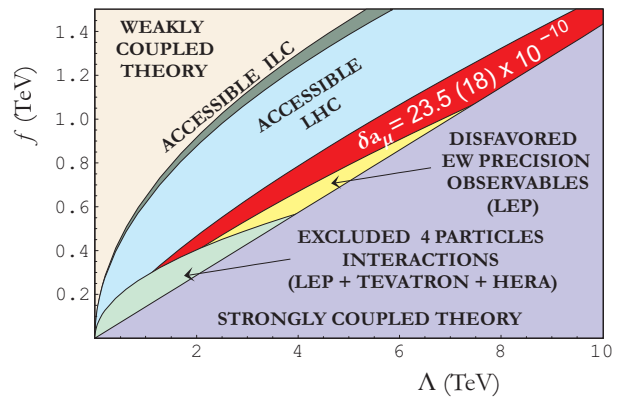}}
\caption {\footnotesize Main limits from branon radiative 
corrections in the $f-\Lambda$ plane for a model with $N=1$.
The (red) central area shows the region in which the branons account for 
the 
muon magnetic moment deficit observed by the E821 Collaboration 
\cite{BNL,gm2}, and
at the same time, are consistent with present collider experiments 
(whose main constraint 
comes from the Bhabha scattering at LEP) and  electroweak precision 
observables. Prospects for future colliders are also plotted.} \label{rad}
\end{center}
\end{figure}

\begin{figure}[bt]
\begin{center}
\resizebox{8.8cm}{6.4cm} 
{\includegraphics{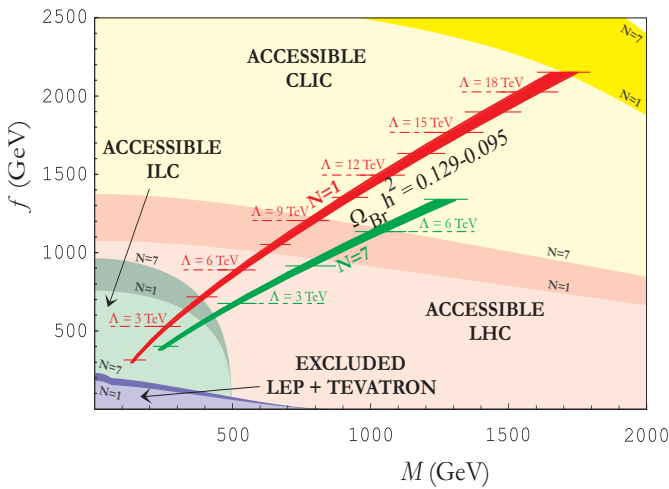}}
\caption {\footnotesize Branon abundance in the range:
$\Omega_{Br}h^2=0.129 - 0.095$,  
 in the $f-M$ 
plane (see \cite{CDM1} for details). The regions are only plotted
for the preferred 
values of the brane tension scale $f$.
One can find on these lines the central values of $\Lambda$ 
corresponding to the observed difference between
the experimental and the SM prediction on the muon anomalous 
magnetic moment 
\cite{BNL,gm2}. 
The lower area is excluded by single-photon processes at LEP-II
together with monojet signals at Tevatron-I
\cite{ACDM}. Prospects for the sensitivity in collider 
searches of real branon production 
are plotted also for future experiments (See \cite{ACDM} and Table \ref{tabHad}).
In this figure one can observe explicitly the dependence on the number of branons $N$,
 since all these regions are plotted for the extreme values $N=1$ and $N=7$.
%
} 
\label{CDM}
\end{center}
\end{figure}

Indeed, if there
 is new physics in the muon anomalous
magnetic moment and it is due to branon radiative
 corrections, the phenomenology
of these particles should be observed at the LHC
and in a possible future ILC, which have  
larger sensitivities for virtual effects working at
a  center of mass energy of $1$ TeV (in contrast with the
direct branon production, where the LHC presents a larger 
sensitivity in any case,
see Tables \ref{tabHad} and \ref{radcoll}, and Figures 
\ref{rad} and \ref{CDM}). In particular,
the LHC should observe an important difference with respect to the 
SM prediction
in channels like $pp\rightarrow e^+e^-$.
The ILC should observe the most important effect in
the Bhabha scattering.

Another limitation to the branon parameters could be obtained  
from electroweak precision measurements, which use to be very 
useful to constrain
models of new physics. The so called oblique corrections
(the ones corresponding to the $W$, $Z$ and $\gamma$ two-point
functions) use to be described in terms of the $S,T,U$ \cite{STU}
or the $\epsilon_1,\epsilon_2$ and $\epsilon_3$ parameters
\cite{Alta}. The experimental values obtained by LEP \cite{EWWG,unknown:2004qh}
are consistent with the SM prediction for a light Higgs 
$m_H\leq 237$ GeV at 95 \% c.l.
In principle, it is necessary to know this parameter in order 
to put constraints on new physics,
but one can talk about disfavored regions of parameters in order 
to avoid fine tunings. 
We can estimate this area by performing a computation of the parameter 
$\bar\epsilon\equiv \delta M_W^2/M_W^2 - \delta M_Z^2/M_Z^2$,
 in a similar way as it was done for the first order correction coming from the
Kaluza-Klein gravitons in the ADD models for rigid branes \cite{CPRS}. 
The experimental value of $\bar\epsilon$ obtained from LEP \cite{EWWG,unknown:2004qh}
is $\bar\epsilon=(1.27 \pm 0.16)\times 10^{-2}$. The theoretical 
uncertainties are one order of magnitude smaller
\cite{Alta} and therefore, we can estimate the constraints for 
the branon contribution
at 95 \% c.l. as $|\delta\bar\epsilon|\,\lsim\, 3.2\times 10^{-3}$ with 
the result \cite{radcorr}:
\begin{equation}\label{eb}
    \frac{f^4}{ N^{1/2}\Lambda^3}\;\gsim\;3.1\; 
\mbox{GeV}\; (\,95\; \%\; c.l.\,)
\end{equation}

This bound has a different dependence on $\Lambda$ than the
SM interactions induced by virtual branons. The constraints coming from this analysis are
complementary to the previous ones. In Figure \ref{rad}, we have included
all the limits in the $f-\Lambda$ plane from 
virtual branon effects. We have also plotted the region in which the effective theory
can be considered as strongly coupled, ($\Lambda\,\gsim\,4\sqrt{\pi}fN^{-1/4}$) and  for which
the loop expansion is no longer valid \cite{radcorr}. We see that the  region compatible
with the Brookhaven results extends for $1100\, \lsim \,
\Lambda\,\lsim\, 15100\, N^{-1/2}$ 
(GeV) and $300\, N^{1/8}\,\lsim\, f\,\lsim\,  2130\, N^{-1/4}$ (GeV).

It is remarkable to note that the same 
parameter space which explains the magnetic 
moment deficit of the muon,  is able to explain the
DM content of the Universe and, in addition,  
the preferred scale is related with the electroweak sector. 
In fact, if the branon mass is of the order of the electroweak scale, 
or more precisely, it  is between $M\sim 100$ GeV and $M\sim 1.7$ TeV, 
then branons could form the total
non baryonic DM abundance observed by WMAP \cite{CDM,CMBR}. 
In Figure \ref{CDM}, we have plotted the $f-M$  regions in which
branons could explain the WMAP measurements. 
We include also the limits from colliders and the 
values of $\Lambda$ corresponding to the central values of  the 
muon anomalous magnetic
 moment observed at Brookhaven. In these regions
branons decouple at 
$T<M<f<\Lambda$, i.e., 
 they are non-relativistic,  behave as cold DM 
and  the effective
theory described by the Lagrangian ($\ref{lag}$) 
can be used to properly  evaluate their thermal relic abundance 
\cite{CDM}.

To summarize, we have shown that
massive branons could offer an alternative explanation 
for the observed dark matter abundance and the recent 
measurements of the muon anomalous magnetic moment. The model only
contains three parameters and, in the region compatible 
with the experiments, 
their values satisfy the natural hierarchy of a weakly coupled
low-energy effective theory i.e. $M\,\lsim\,f\,\lsim\,\Lambda$. We have also shown that
in the preferred parameter region, future colliders such as 
LHC or ILC should 
be sensitive to branon phenomenology.

{\em Acknowledgments} --- 
This work is partially supported by DGICYT (Spain) under project numbers
FPA 2004-02602 and BFM 2002-01003, by NSF grant No.~PHY--0239817
and by the Fulbright-MEC (Spain) program.


\vspace{-.6cm}

\begin{thebibliography}{999}

\bibitem{CMBR}
  D.N. Spergel {\it et al.}, 
  {\it Astrophys. J. Suppl.} {\bf 148}, 175 (2003) 

\bibitem{SNeIa}
  S. Perlmutter {\it et al.}, 
   {\it Astrophys. J.} {\bf 517}, 565 (1999); 
  A.G. Riess {\it et al.}, 
{\it Astrophys. J.}, {\bf 607}, 665 (2004) 

\bibitem{LSS}
  A.C. Pope {\it et al.}, 
  {\it Astrophys. J.} {\bf 607}, 655 (2004)
  
\bibitem{Lyalpha}
  R.A.C. Croft {\it et al.},
  {\it Astrophys. J.}  {\bf 581}, 20 (2002)
 

\bibitem{neutralino}
G. Jungman, M. Kamionkowski 
and K. Griest, {\it Phys. Rep.} {\bf 267}, 195 (1996)

\bibitem{PDG}
S. Eidelman {\it et al.}, Phys. Lett. B {\bf 592}, 1 (2004)


\bibitem{BNL}
H.N.~Brown {\it et al.}, 
{\it Phys.\ Rev.\ Lett.} {\bf 86}, 2227 (2001);
G.W.~Bennett {\it et al.}, 
{\it Phys.\ Rev.\ Lett.} {\bf 89}, 101804 (2002)
and 
{\it Phys.\ Rev.\ Lett.} {\bf 92}, 161802 (2004)

\bibitem{gm2}
  M.~Passera,
  hep-ph/0411168;
  J.F.~de Troconiz and F.J.~Yndurain,
  hep-ph/0402285;
  A.~Hocker,
  hep-ph/0410081


\bibitem{Baltz:2001ts}
  E.A.~Baltz and P.~Gondolo, {\it Phys. Rev. Lett.}
{\bf 86}, 5004 (2001)

\bibitem{ADD} N. Arkani-Hamed, S. Dimopoulos and G. Dvali,
{\it Phys. Lett.} {\bf B429}, 263 (1998) and {\it Phys. Rev.} {\bf D59}, 086004 (1999); I. Antoniadis {\it et al.},
{\it Phys. Lett.} {\bf  B436} 257  (1998)

\bibitem{CDM} J.A.R. Cembranos, A. Dobado and A.L. Maroto, {\it
Phys. Rev. Lett.} {\bf 90}, 241301 (2003)

\bibitem{M} 
A.L. Maroto, {\it Phys. Rev.} {\bf D69}, 043509
(2004) and {\it Phys. Rev.} {\bf D69}, 101304 (2004);
J.A.R. Cembranos, A. Dobado and A.L. Maroto   {\it Int. J. Mod. Phys. } {\bf D13}, 2275 (2004)




\bibitem{DoMa} R. Sundrum, {\it Phys. Rev.} {\bf D59}, 085009 (1999);
A. Dobado and A.L. Maroto {\it Nucl. Phys.} {\bf B592}, 203 (2001)

\bibitem{BSky} J.A.R. Cembranos, A. Dobado and A.L. Maroto,
{\it  Phys. Rev.} {\bf D65}, 026005 (2002) 

\bibitem{ACDM} J. Alcaraz {\it et al.}, {\it Phys. Rev.} {\bf D67}, 075010
(2003); J.A.R. Cembranos, A. Dobado, A.L. Maroto,
  {\it Phys. Rev.} {\bf D70}, 096001 (2004) 
   and {\it AIP Conf. Proc.} {\bf 670}, 235 (2003);
P. Achard {\it et al.}, 
{\it Phys. Lett.} {\bf B597}, 145 (2004)

\bibitem{CrSt} P. Creminelli and A. Strumia,
{\it Nucl. Phys.} {\bf B596}, 125 (2001)

\bibitem{CDM1}
J.A.R. Cembranos, A. Dobado and A.L. Maroto 
{\it Phys. Rev.} {\bf D68}, 103505 (2003) 




\bibitem{radcorr}
J.A.R. Cembranos, A. Dobado and A.L. Maroto, in preparation 

\bibitem{branemuon}
M.L.~Graesser, {\it Phys.\ Rev.}  {\bf D61}, 074019 (2000);
%
  S.C.~Park and H.S.~Song,
  {\it Phys.\ Lett. } {\bf B523}, 161 (2001);
%
  K.~Sawa,
  hep-ph/0506190.

\bibitem{CPRS}  R. Contino, L. Pilo, R. Rattazzi and
A. Strumia {\it JHEP} {\bf 0106}, 005 (2001)

\bibitem{GS}  G. Giudice and A. Strumia, {\it Nucl. Phys.} {\bf B663}, 377 (2003)

\bibitem{Hewett}
J.L. Hewett, {\it Phys.\ Rev.\ Lett.} {\bf 82}, 4765 (1999)

\bibitem{Adloff:2003jm}
C.~Adloff {\it et al.}, 
{\it Phys. Lett.} {\bf B568}, 35 (2003)

\bibitem{d0} B. Abbott {\it et al.}, 
 {\it Phys.\ Rev.\ Lett.} {\bf 86}, 1156 (2001)

\bibitem{unknown:2004qh}
D. Abbaneo {\it et al.}, 
hep-ex/0412015

\bibitem{STU}
M.E. Peskin, T. Takeuchi, {\it Phys.\ Rev.} {\bf D46}, 381 (1992)

\bibitem{Alta}
  G.~Altarelli, R.~Barbieri and F.~Caravaglios,
  {\it Int.\ J.\ Mod.\ Phys.\ } {\bf A13}, 1031 (1998)

\bibitem{EWWG} 
G.~Altarelli, hep-ph/0406270



\end{thebibliography}
\end{document}